\documentclass[prd,reprint,amsmath,amssymb,aps,nofootinbib]{revtex4-2}

\usepackage{graphicx}
\usepackage{dcolumn}
\usepackage{bm}

\begin{document}

\preprint{APS/123-QED}

\title{The phenomenology of Quantum Bounce \\in the Klein-Gordon, Wheeler-DeWitt and Dirac formalisms}

\author{Eleonora Giovannetti}
\email{eleonora.giovannetti@uniroma1.it}
\affiliation{Department of Physics, La Sapienza University of Rome, P.le Aldo Moro, 5 (00185) Roma, Italy}

\date{\today}

\begin{abstract}
We investigate the different meanings that the concept of Quantum Bounce acquires in various formalisms. The original idea refers to the phenomenology that appears in the Klein-Gordon framework when homogeneous cosmologies are considered. In that case, the Quantum Bounce describes the quantum scattering between a collapsing and an expanding Universe branch, and therefore provides a quantum description of the semiclassical Big Bounce mechanism. Here, we show that the proposal of the Quantum Big Bounce is well-grounded, thanks to the computation of the volume operator mean values and its standard deviation in the Wheeler-DeWitt framework for the isotropic case. Then, we analyze the Bianchi models in the Dirac approach, now showing that the Quantum Bounce concept can be implemented to describe the Kasner transitions of the Belinski–Khalatnikov–Lifshitz map at a quantum level. In summary, the quantum scattering framework borrowed from particle physics can serve as a good model for different cosmological scenarios, which can exhibit scalar-like or fermionic-like behaviours depending on how the anisotropies are described in the dynamics. 
\end{abstract}

\maketitle

\section{Introduction}
As demonstrated in the renowned Singularity Theorems by Hawking and Penrose \cite{PH1,PH2}, the presence of spacetime singularities is a very general property of the Einstein equations. This can be interpreted as a sign that General Relativity loses its predictability near the Planckian region, where quantum effects start to play a relevant role. Hence, a quantum theory of gravity is required in order to solve the non-physical predictions of the Einstein theory that emerge when it is applied out of the validity domain. Loop Quantum Gravity \cite{ThiemannBook,CQG} represents one of the most reliable Quantum Theories of Gravity, whose main result is the emergence of a discrete spectrum for the area and volume geometrical operators \cite{RS}. In the cosmological sector of the theory, i.e. Loop Quantum Cosmology, this feature produces deviations from General Relativity at a semiclassical level, giving birth to the most successful scenario in which the Big Bang singularity is solved, that is known as Big Bounce. This mechanism of regularization replaces the initial singularity with a symmetrical reconnection of the contracting and expanding branches of the Universe evolution in correspondence of a minimum volume \cite{Ashtekar2006,AshtekarI,Ashtekar2008,Ashtekar2009,Ashtekar2011,Review}. However, there are still some open questions in Loop Quantum Gravity and Loop Quantum Cosmology, especially regarding the correct procedure through which transpose the constraints from the full quantum picture to the reduced cosmological sector \cite{ThiemannBook,Bojowald,Bruno}. Indeed, Loop Quantum Cosmology is constructed by imposing the cosmological symmetries before quantizing. 

In \cite{QB1,QB2}, we propose a new treatment of the Big Bounce in order to go beyond the pure semiclassical picture. Indeed, when the Universe is in the very primordial phase of its evolution, i.e. at a Planckian energy density, the semiclassical dynamics cannot capture all the features of the full quantum state. Therefore, we introduce a new concept of Big Bounce as a pure quantum phenomenon related to the interaction between the two separated branches of the Universe evolution, when a time-dependent potential is present. Actually, the analogy with the relativistic scattering processes is clear when homogeneous cosmological models are considered. In particular, in \cite{QB1} the Bianchi I cosmology with an ekpyrotic-like potential is studied by choosing the isotropic Misner variable as time after quantization. Instead, in \cite{QB2} the Friedmann-Lemaître-Robertson-Walker (FLRW) Universe is investigated with a self-interacting scalar field and a matter relational time \cite{Rovelli}. The results presented in \cite{QB1,QB2} show that the probability density associated to this quantum transition has the phenomenology of a bouncing cosmology in the semiclassical limit. However, its presence is not due to the existence of a minimum volume in the semiclassical dynamics, but only to the time-dependent potential that acts as a quantum interacting term and then mixes the two Universe branches. This means that a bouncing picture can be recovered at the level of the Wheeler-DeWitt equation \cite{deWitt1} thanks to pure quantum effects. Also, the interesting aspect of this approach is that the resolution of the initial singularity does not rely on the semiclassical concept of regularized trajectories, and in fact the Universe is still singular at a classical level. 

The aim of this paper is to make stronger the Quantum Big Bounce prediction by looking at the behaviour of the Universe volume operator when a relational time is considered, in order to avoid the misleading interpretation of the isotropic Misner variable as a monotonic time variable in a bouncing scenario. Also, we want to show the multiple applications of this Quantum Bounce approach by analyzing the Bianchi models in the Dirac formalism. In this case, the quantum scattering is not related to the initial Big Bounce but to the Kasner transitions of the Belinski–Khalatnikov–Lifshitz (BKL) map, i.e. the reflections of the point Universe off the potential during the dynamics towards the initial singularity. We recall that a quantum version of the BKL map is still lacking, so the Quantum Bounce approach could lead to interesting implications in this framework. Finally, this paper shows that the same concept can be revised in order to describe different cosmological phenomenologies. In particular, the Quantum Bounce approach represents a step closer to the goal of providing a full quantum description of the primordial Universe dynamics.

The paper is structured as follows. In Section \ref{KG} we review the Quantum Big Bounce formalism introduced in \cite{QB2} in which the isotropic Universe is studied in view of the Klein-Gordon formalism by using a relational time approach. Then, in Section \ref{WDW} we compare the behaviour of the Universe volume operator in the Wheeler-DeWitt formalism with the Quantum Big Bounce picture. In Section \ref{D} we promote the Klein-Gordon formalism to the Dirac one and we explain why this framework is useful to study the BKL map at a pure quantum level. Finally, in Section \ref{Conc} we conclude the paper by outlining a summary and discussing the results.   

We will use natural units $\hslash=c=8\pi G=1$.

\section{The Klein-Gordon formalism:\\the Big Bounce as a quantum scattering\label{KG}}
In this Section, we review the Quantum Big Bounce approach as first presented in \cite{QB1,QB2}, where we introduced a quantum probabilistic interpretation of the Big Bounce by setting up an analogy between the expanding/collapsing solutions of the Wheeler-DeWitt equation and the positive/negative frequency solutions of the corresponding Klein-Gordon equation. 
In this picture, the Big Bounce is a quantum transition between a collapsing and an expanding Universe in the presence of an interaction term. The associated probability density reproduces the typical symmetrical reconnection of the semiclassical Big Bounce, hence the interpretation as \emph{Quantum} Big Bounce. 

The formalism we use is the relativistic scattering theory in the Klein-Gordon formulation \cite{BjD}. In particular, in the presence of a time-dependent potential, the solution to the Klein-Gordon equation
\begin{equation}
(\Box_x + m^2 + U(x)) \Psi(x,t) = 0
\end{equation}
can be written in terms of the propagator $\Delta_F(x-y)$ for the free scalar field $\Phi(x,t)$, i.e.
\begin{equation}
\label{propagator}
\Psi(x,t) = \psi(x,t) + \int \Delta_F(x-y) U(y) \Psi(y,t)\,d^3 y \,.
\end{equation}
Hence, the positive frequency solutions (i.e. particles) move forward in time and the negative frequency ones (i.e. anti-particles) backward. In this wavefunction picture, the transition amplitude between two single-particle states is the projection of the interacting solution \eqref{propagator} onto the initial free one $\psi(x,t)$ by means of the Klein-Gordon scalar product. Using \eqref{propagator}, it can be written as\footnote{In this definition we are considering only the cases of pair production and annihilation.}
\begin{equation}
S=-i\int d^3y\,\psi^*(y)U(y)\Psi(y)\,.	\label{Smatrix}    
\end{equation}
Then, the probability associated to the scattering is simply the square modulus of the amplitude \eqref{Smatrix}.

In the following we will present only the case of \cite{QB2}, in which the Big Bounce is studied for the isotropic Universe in the relational time approach. In this way we avoid the non-trivial interpretations that emerge when a volume-like time variable is adopted, as in \cite{QB1}.

\subsection{The isotropic case}
The Wheeler-DeWitt equation for the FLRW model with the normal ordering convention is
\begin{equation} \label{WDW1}
\hat{\mathcal{H}}_\text{FLRW}\Psi(\alpha,\phi)=(\partial_\alpha^2-\partial_\phi^2)\Psi(\alpha,\phi)= 0\,,
\end{equation}
in which it clearly appears the formal analogy with a (1+1)-dimensional massless Klein-Gordon equation for a relativistic particle. In particular, we can identify $\alpha$ (i.e. the variable related to the Universe volume $V=e^{3\alpha}$) as the spatial degree of freedom, and $\phi$ as a relational time variable after quantization \cite{Rovelli}. As is well known, the most general solution is a linear combination of plane waves
\begin{equation}
\label{plane}
\varphi^{\pm}(\alpha,\phi)=e^{i(k\alpha\mp \omega_k\phi)}\,,  
\end{equation}
where $k$ is the wavenumber and $\omega_k = |k| $ is the frequency. Moreover, in the hypothesis of highly-localized wavepackets, it is possible to evaluate the quantum operators $\hat{p}_\alpha=-i\partial_\alpha$ and $\hat{p}_\phi=-i\partial_\phi$ along the classical trajectories. In particular, by considering the equation of motion
\begin{equation}
\label{rel}
\frac{d\alpha}{d\phi} = -\frac{p_\alpha}{p_\phi}\,,
\end{equation}
that has a covariant character, we derive that the relative sign between the constant of motion $p_\alpha$ and $p_\phi$ fixes on which branch of the evolution the Universe evolves: it collapses if the ratio is positive, it expands if it is negative. Therefore, far from the full quantum region of the initial singularity, we can associate positive energy states (i.e. particles states) 
\begin{equation}
\label{psi+}
\psi^{+}(\alpha,\phi) =\int_{-\infty}^{+\infty}dk\,\frac{e^{-\frac{(k-\overline{k}_+)^2}{2\sigma_+^2}}e^{i(k\alpha- \omega_k\phi)}}{\sqrt{2\pi \sigma^2_+}}
\end{equation}
to the expanding Universe, and negative energy states (i.e. anti-particles states) 
\begin{equation}
\label{psi-}
\psi^{-}(\alpha,\phi) =\int_{-\infty}^{+\infty} dk\,\frac{e^{-\frac{(k-\overline{k}_-)^2}{2\sigma_-^2}}e^{i(k\alpha+ \omega_k\phi)}}{\sqrt{2\pi \sigma^2_-}}
\end{equation}
to the collapsing Universe.  

\begin{figure}[h!]
\centering
\includegraphics[scale=0.31]{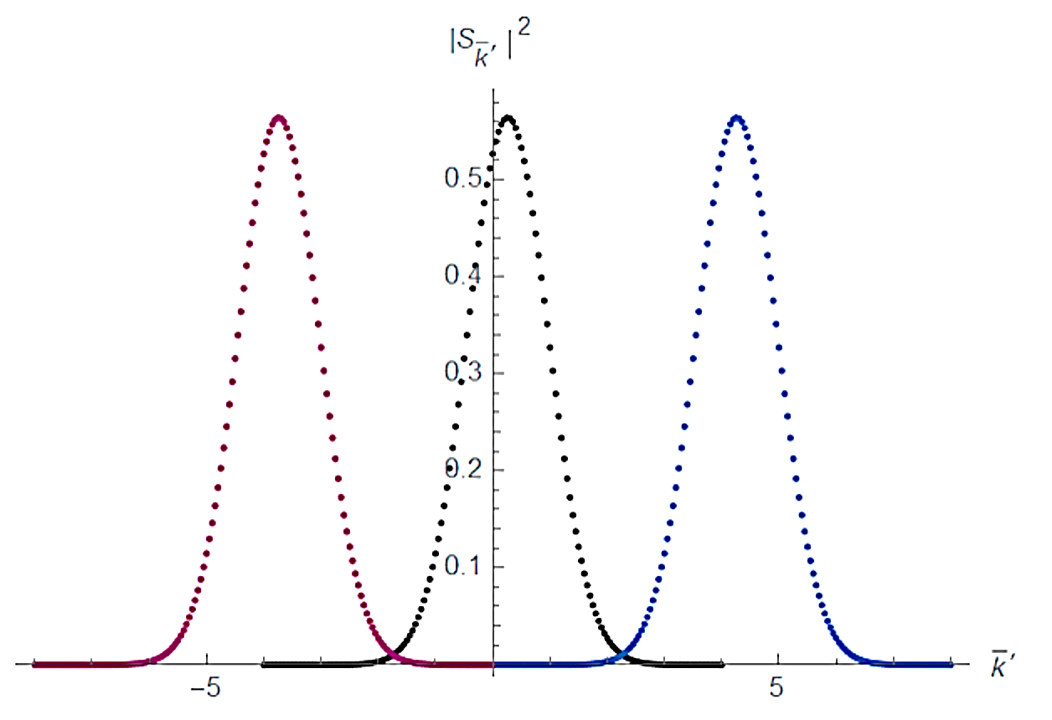}
\caption{Plot of $|S(\bar{k'},\bar{k})|^2$ as function of $ \bar{k}'$, where $\bar{k},\bar{k}'$ are respectively the wavenumbers of the initial and final state of the Universe. From the left, $\bar{k}=-4,0,4$. Credits: \cite{QB2}.}
\label{fig:ampiezzaI}
\end{figure}

In the framework traced above, a time-dependent potential is the only missing ingredient to make possible the interaction between the two separated branches of the Universe evolution. In this case, the Wheeler-DeWitt equation \eqref{WDW1} turns out to be
\begin{equation}
\label{WDWU}
\left[\partial^2_\alpha  - \partial^2_\phi + U(\phi)\right]\Psi(\alpha,\phi) = 0\,,
\end{equation}
and the Quantum Big Bounce scattering amplitude simply the Klein-Gordon projection of the interacting solution $\Psi(\alpha,\phi)$ onto the initial free one $\psi^-(\alpha,\phi)$, as presented in \eqref{Smatrix}. Figure \ref{fig:ampiezzaI} shows the Quantum Big Bounce probability density $|S|^2$ for the isotropic Universe when $U(\phi)$ is an ekpyrotic potential (see \cite{QB2} for all the derivation).

The most relevant result of the analysis performed in \cite{QB1,QB2} is that the probability density is always well defined, thus showing that a Quantum Big Bounce is always possible. Moreover, when the initial and final states are highly localized, the most likely transition is the one that preserves the quantum numbers from the initial to the final Universe state, thus reproducing the same phenomenology of the semiclassical bouncing picture.

In the next Section we will compare this probabilistic approach with the behaviour of the Universe volume operator, searching for evidences of the Quantum Big Bounce in the Wheeler-DeWitt formalism. 
\section{The Wheeler-DeWitt formalism:\\imprints of the Quantum Big Bounce\\ on the Universe volume operator\label{WDW}}
In this Section, we are interested in verifying if there are imprints of the Quantum Big Bounce process on the spectrum of the Universe volume operator $\hat{V}$. Actually, when dealing with an exact solution for the interacting wavefunction, it is possible to compute the $\hat{V}$ eigenvalues on it, and, where relevant, the variances in function of the relational time variable $\phi$.

Here, we consider the quantum interacting term $U(\phi)= \lambda/\phi^2$, with $\lambda > 0$, i.e. a localized potential that is relevant only nearby $\phi=0$. In this way, it is easier to demonstrate the unitarity of the scattering matrix $S$, since the interacting solution reduces to the free plane waves for $\phi\rightarrow\pm\infty$. Therefore, the Wheeler-DeWitt equation \eqref{WDW1} is
\begin{equation}
\label{WDWKG}
\left[\partial^2_\alpha  - \partial^2_\phi + \frac{\lambda}{\phi^2}\right]\Psi(\alpha,\phi) = 0\,,
\end{equation}
whose general solution can be written as
\begin{equation}
\label{Psi}
\Psi(\alpha,\phi) =\frac{1}{\sqrt{2\pi \sigma^2}} \int_{-\infty}^{+\infty} dk\,e^{-\frac{(k-\bar{k})^2}{2\sigma^2}} \Phi_k(\phi)e^{ik\alpha}\,,
\end{equation}
where
\begin{align}
\nonumber
\Phi_k(\phi) =&c_1\sqrt{|\phi|}\mathcal{B}_J\bigg(\frac{1}{2}\sqrt{1+4\lambda},k\phi\bigg)+\\+&c_2\sqrt{|\phi|}\mathcal{B}_Y\bigg(\frac{1}{2}\sqrt{1+4\lambda},k\phi\bigg)\,,
\label{solKG}
\end{align}	
$c_1,c_2$ are constants and $\mathcal{B}_{J},\mathcal{B}_{Y}$ are the Bessel functions of the first and second kind respectively. In particular, for $\lambda=2$ the solution \eqref{solKG} reduces to
\begin{align}
\nonumber
\Phi_k(\phi) =&c_1\sqrt{\frac{2}{\pi k}}\bigg(-\cos{(k\phi)}+\frac{\sin{(k\phi)}}{k\phi}\bigg)+\\+&c_2\sqrt{\frac{2}{\pi k}}\bigg(-\frac{\cos{(k\phi)}}{k\phi}-\sin{(k\phi)}\bigg)\,,
\label{solKG2}
\end{align}
which is a manageable analytic expression for the next computations. We remind that the initial singularity is still present at a classical level and the Universe evolves along one of the two singular branches depending on the initial conditions on the motion. At a quantum level, the presence of the time-dependent potential prevents the frequency separation, making the Universe wavefunction \eqref{solKG} necessarily a mixed state that reduces to the free solutions in the limit $\lambda=0$. In this particular case, the same limit is reached for $\phi\rightarrow\pm\infty$. Moreover, since the potential is pretty steep, the deviation from the free case is significant only in the nearby of $\phi=0$. This fact suggests that a similar modified behaviour is not sufficient to regularize the quantum behaviour of the volume operator $\hat{V}$ and, more in general, that the Wheeler-DeWitt formalism is not the right tool to solve the initial singularity, as we will show at the end of this Section. 

\begin{figure}[ht]
\centering
\includegraphics[scale=0.6]{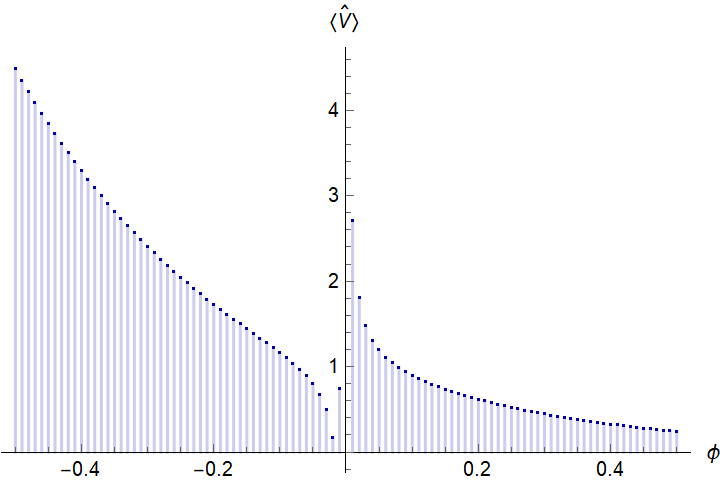}
\includegraphics[scale=0.6]{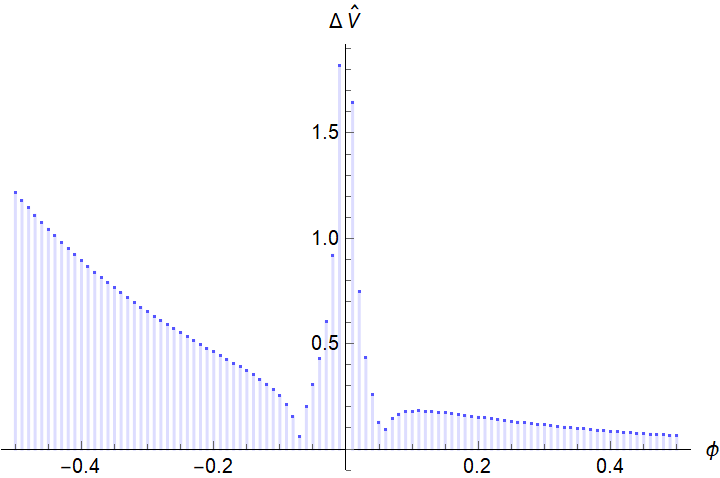}
\caption{Plot of $\langle\hat{V}\rangle$ and $\Delta\hat{V}$ in function of $\phi$ computed on $\Psi(\alpha,\phi)$.}
\label{V}
\end{figure}

\begin{figure}[ht]
\centering
\includegraphics[scale=0.68]{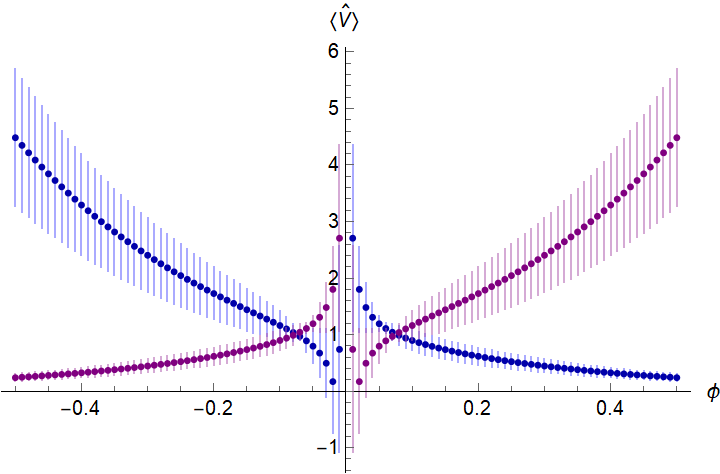}
\caption{Plot of $\langle\hat{V}\rangle$ with error bars $\Delta\hat{V}$ in function of $\phi$, computed on $\Psi(\alpha,\phi)$ (blue trajectory) and $\Psi^*(\alpha,\phi)$ (purple trajectory).}
\label{VdeltaV}
\end{figure}

The initial conditions on \eqref{solKG2} are fixed by imposing that $\Phi_k(\phi)\rightarrow e^{ik\phi}$ for $\phi\rightarrow-\infty$, i.e. a collapsing state, thus obtaining  
\begin{equation}
    \Phi_k(\phi)=\frac{e^{ik\phi}(i+k\phi)}{k\phi}\,.
\end{equation}
Then, we construct a Gaussian wavepacket
\begin{equation}
    \Psi(\alpha,\phi)=\int_{-\infty}^{+\infty} dk\,e^{-(k-\bar{k})^2/\sigma^2}e^{ik\alpha}\Phi_k(\phi)\,,
    \label{Psi}
\end{equation}
in order to study the quantum behaviour of the Universe volume operator. In this representation, the volume operator acts as $\hat{V}=e^{3\alpha}$, and the meanvalues as a function of $\phi$ can be computed as
\begin{equation}
    \langle\hat{V}\rangle (\phi)=\frac{\displaystyle\int_{-\infty}^{+\infty}d\alpha\,[\Psi^*\partial_\phi(\hat{V}\Psi)-(\hat{V}\Psi)\partial_\phi\Psi^*]}{\displaystyle\int_{-\infty}^{+\infty}d\alpha\,[\Psi^*\partial_\phi\Psi-\Psi\partial_\phi\Psi^*]}\,.
\end{equation}
In Figure \ref{V} we can see that $\langle\hat{V}\rangle$ seems to reproduce a collapsing branch for the Universe quantum evolution, with a deviation around $\phi=0$ in which the interacting potential is dominant. Moreover, in that region the standard deviation $\Delta\hat{V}=\sqrt{\langle\hat{V}^2\rangle-\langle\hat{V}\rangle^2}$ grows rapidly, becoming of the same order as $\hat{V}$.
This means that the wavepacket $\Psi(\alpha,\phi)$ stops following the semiclassical collapsing trajectory and that the quantum meanvalues are no more predictive about the Universe evolution in that region, no matter how $\Psi(\alpha,\phi)$ is localized. Indeed, from Figure \ref{VdeltaV} we can clearly see that in the region around $\phi=0$ the significant growth of the standard deviation $\Delta\hat{V}$ makes the meanvalue of $\hat{V}$ compatible with a semiclassical expanding trajectory in the positive semiaxis, which is obtained by computing $\langle\hat{V}\rangle$ on 
\begin{equation}
\Phi^*_k(\phi)=\frac{e^{-ik\phi}(-i+k\phi)}{k\phi}\,,
\end{equation}
that reduces to $e^{-ik\phi}$ for $\phi\rightarrow-\infty$, thus representing an expanding state.
This evidence can be interpreted as a semiclassical trace of the Quantum Big Bounce transition at the level of the Wheeler-DeWitt. 
\begin{figure}[ht]
\centering
\includegraphics[scale=0.7]{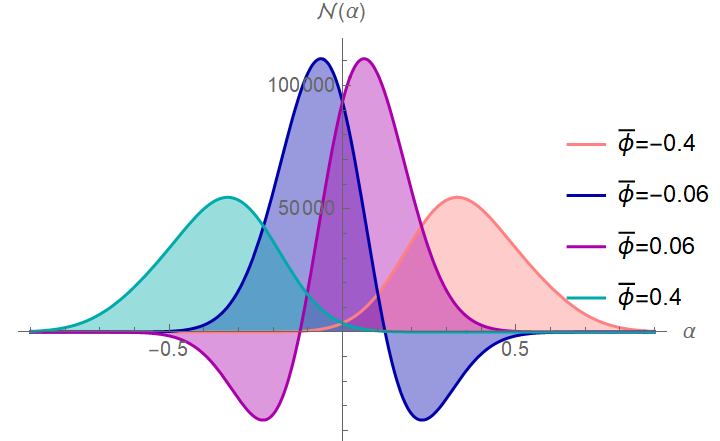}
\caption{Plot of the Klein-Gordon density $\mathcal{N}(\alpha)$ associated to the wavepacket $\Psi(\alpha,\phi)$ for different values of time.}
\label{norm}
\end{figure}
Indeed, in Figure \ref{norm} we can see that the Klein-Gordon density
\begin{equation}
\mathcal{N}(\alpha)=[\Psi^*(\alpha,\phi)\partial_\phi\Psi(\alpha,\phi)-\Psi(\alpha,\phi)\partial_\phi\Psi^*(\alpha,\phi)]|_{\phi=\bar{\phi}}    
\end{equation}
is positive-defined only in the region in which the potential is negligible. On the other hand, in the vicinity of $\phi=0$, negative frequencies start to contribute to $\Psi(\alpha,\phi)$, proving that the deviation of the meanvalues trajectory is not an artifact but it is due to the emergence of expanding-like states in the wavepacket distribution because of the interaction with the time-dependent potential. Although, we also demonstrated that the Klein-Gordon norm is conserved, hence the unitarity of the process is ensured.

\begin{figure}
    \centering
\includegraphics[width=0.9\linewidth]{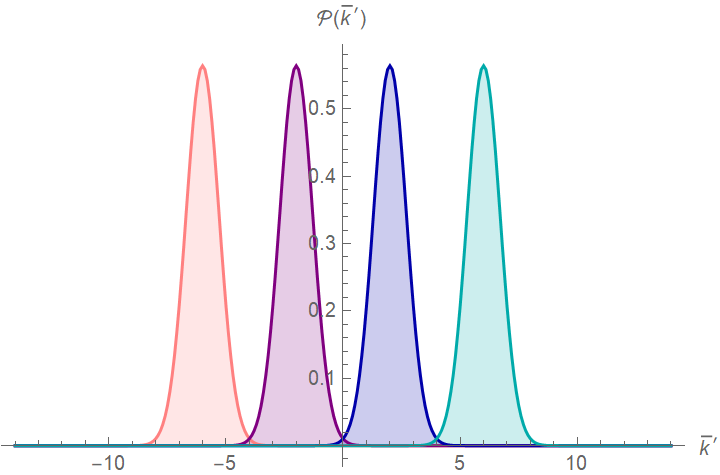}
    \caption{Quantum Big Bounce probability density $\mathcal{P}_{\bar{k}'}$ in the presence of the potential $U(\phi)=2/(\phi^2+\epsilon)$, with $\epsilon=10^{-100}$. From the left, we fix $\bar{k}=-6,-2,2,6$.}
    \label{Dio}
\end{figure}

This analysis highlights the limits of this approach in solving the singularity by simply looking at the meanvalues of the quantum operators. On the other hand, the scattering amplitude formalism, as presented in \cite{BjD} and developed for the isotropic Universe in Section \ref{KG}, is the natural approach to adopt in this case, since it returns the probability associated with the quantum transition from the contrancting branch to the expanding one. Moreover, the propagator formalism implements the correct boundary conditions by construction, making it possible the emergence of the expanding Universe far away from the interaction with the time-dependent potential, independently from the free initial state of the Universe evolution (here, a contracting one). In particular, by making explicit  the expression of the Feynman propator $\Delta_F(x-y)$, we get
\begin{widetext}
\begin{align}
\Psi(\alpha,\phi)=\psi(\alpha,\phi)-i\int dk\,\varphi_k^+
(\alpha,\phi)\int d\alpha'd\phi'\theta(\phi-\phi'){\varphi_k^+}^*(\alpha',\phi')U(\phi')\Psi(\alpha',\phi')\nonumber\\-i\int dk\,\varphi_k^-
(\alpha,\phi)\int d\alpha'd\phi'\theta(\phi'-\phi){\varphi_k^-}^*(\alpha',\phi')U(\phi')\Psi(\alpha',\phi')\,,
\label{Psiprop}
\end{align}
\end{widetext}
and then we can write the Quantum Big Bounce scattering amplitude as
\begin{align}
S_{k',k}=\lim_{t\rightarrow +\infty}\int d\alpha\,{\psi_{k'}^+}^*(\alpha,\phi)i\overleftrightarrow{\partial_\phi}\Psi_k(\alpha,\phi)=\\=-i\int d\alpha\,d\phi\, {\psi_{k'}^+}^*(\alpha,\phi)U(\phi)\Psi_k(\alpha,\phi)\,,
\label{SBounce}
\end{align}
thus obtaining the same expression as in \eqref{Smatrix}.
Our information on the analytical solution of \eqref{WDWKG} makes the expression \eqref{Psiprop} exact, although this approach is perturbative and is valid at all orders of expansions of $\Psi(\alpha,\phi)$. Now, we can compute the probability associated to the Quantum Big Bounce transition by taking the square modulus of \eqref{SBounce}, in which $\psi^+_{k'}(\alpha,\phi)$ is a free expanding wavepacket (see \eqref{psi+}) and $\Psi_k(\alpha,\phi)$ is an interacting collapsing-like wavepacket (see \eqref{Psi}). In order to make the integral in $\phi$ well defined on the entire real line, we consider $U(\phi)=2/(\phi^2+\epsilon)$ taking $\epsilon\ll1$. Therefore, the probability density takes the expression
\begin{equation}
\mathcal{P}_{\bar{k}',\bar{k}}^{\epsilon}=\bigg|\int d\phi\,dk\,\frac{2\,e^{-(k-\bar{k}')^2}e^{-(k-\bar{k})^2}e^{ik\phi}\Phi_k(\phi)}{\phi^2+\epsilon}\bigg|^2\,,
\end{equation}
in which we have fixed $\sigma=1$ for both the Gaussian wavepackets. In Figure \ref{Dio} we can see the probability density $\mathcal{P}_{\bar{k}'}$ for fixed values of $\bar{k}$ and $\epsilon=10^{-100}$.
Since we chose quite localized wavepackets, we find again that the most likely transition is the one from a collapsing to an expanding Universe with $\bar{k}=\bar{k}'$, as in Section \ref{KG}. We conclude by reminding that this formalism can be applied for any potential and also in the absence of an exact solution for the Wheeler-DeWitt equation, since \eqref{SBounce} gives a non-zero probability even when $\Psi_k(\alpha,\phi)$ is a free collapsing wavepacket.

In the next section, we will explain how to translate the propagator formalism in the Dirac framework, where new interpretations emerge.

\section{The Dirac formalism: \\the Kasner transition as \\a Quantum Bounce\label{D}}

Let us now improve the previous treatment by both generalizing the chosen cosmological model and formalism. In particular, we consider the anisotropic Bianchi II model whose Hamiltonian constraint is
\begin{equation}
\label{WDWBII}
    \mathcal{H}_\text{BII}=-p_\alpha^2+p_+^2+p_-^2+\delta e^{4(\alpha-2\beta_+)}=0\,,
\end{equation}
in which $\delta$ is a constant and $(p_+,\,p_-)$ are the momenta conjugated to the two anisotropic degrees of freedom $(\beta_+,\,\beta_-)$, whereof only one appears explicitly in \eqref{WDWBII}. For the scope of this analysis, the interesting aspects of this model are twice. The first is that the Bianchi II Universe is the simplest homogeneous but anisotropic model in whose Hamiltonian it is naturally present a potential term coming from the extrinsic curvature \cite{PC}. The second is that, after performing a reduction of the Hamiltonian \eqref{WDWBII}, its dynamics towards the singularity can be modeled as that of a two-dimensional point particle that reflects off the potential wall, and then free moves toward the initial singularity. Therefore, this setting seems suitable for implementing the Dirac approach in order to study the reflection at a quantum level. The relevance of this toy model relies in the possibility to extend this scheme to the Bianchi IX model, and eventually to the generic cosmological solution. Actually, by iterating the Bianchi II reflection scheme, it is possible to reconstruct the Bianchi IX dynamics toward the initial singularity, since in that model the potential is a closed triangular domain. In particular, the resulting reflection map is known as the BKL map (see \cite{MisnerMixmaster,MisnerQC69,PC,G,Giovannetti_2019,Segreto,MixmasterGUP} for detailed analyses in standard and modified formalisms). Moreover, the implementation of the Dirac approach represents an improvement over the previous treatment. Indeed, the normalization of the wavepackets in the Klein-Gordon formalism is non-trivial. However, the other side of the coin is that it is more challenging to obtain an exact solution of the Dirac equation (as was possible in Section \ref{WDW}), depending on the expression of the potential term.     

Starting from the Klein-Gordon equation for the Bianchi II model, i.e.
\begin{equation}
    \partial_\alpha^2-\partial_+^2-\partial_-^2+\delta e^{4(\alpha-2\beta_+)}\psi(\alpha_,\beta_\pm)=0\,,
    \label{BII}
\end{equation}
it is easy to verify that the associated Dirac equation is
\begin{equation}
\big[i\sigma^\mu\partial_\mu-\mathbb{I}\,e^{2(\alpha-2\beta_+)}\big]\psi(\alpha_,\beta_\pm)=0\,,
\end{equation}
in which $\mu=(\alpha,+,-$), $\mathbb{I}$ is the identity and $\sigma_\mu$ are the usual Pauli matrices, i.e.
\begin{align}
\mathbb{I}=
\begin{pmatrix}
1 & 0 \\
0 & 1 \\ 
\end{pmatrix}\,,\quad
\sigma_\alpha=
\begin{pmatrix}
1 & 0 \\
0 & -1 \\ 
\end{pmatrix}\,,
\\
\sigma_+=
\begin{pmatrix}
0 & 1 \\
1 & 0 \\ 
\end{pmatrix}\,,\quad
\sigma_-=
\begin{pmatrix}
0 & -i \\
i & 0 \\ 
\end{pmatrix}\,.
\end{align}
The solution $\psi(\alpha,\beta_\pm)$ describing the Universe behaviour is now a two-components object (for an interesting discussion about the interpretation of the Bianchi I Universe solution as a fermion, see \cite{DiracBianchi}). In particular, the collapsing Universe can be written as
\begin{equation}
    \psi^{-}(\alpha_,\beta_\pm)=\iint_{-\infty}^{+\infty}u(k_\pm)N^-(k_\pm)e^{-i(k\alpha-k_+\beta_+-k_-\beta_-)}dk_\pm\,,
\end{equation} 
and the expanding one as
\begin{equation}
    \psi^{+}(\alpha_,\beta_\pm)=\iint_{-\infty}^{+\infty}v(k_\pm)N^+(k_\pm)e^{-i(k \alpha-k_+\beta_+-k_-\beta_-)}dk_\pm\,,
\end{equation} 
in which $N^\mp(k_\pm)=A^{\mp}e^{-(k_+-\bar{k}_+^{\mp})^2}e^{-(k_--\bar{k}_-^{\mp})^2}$, $k=\sqrt{k_+^2+k_-^2}$ and
\begin{equation}
u(k_\pm)=
    \begin{pmatrix}
        k \\ -(k_++ik_-)
    \end{pmatrix}
   \,, \;
    v(k_\pm)=
    \begin{pmatrix}
        k \\ k_++ik_-
    \end{pmatrix} \,.
\end{equation}
Also, the two Gaussian wavepackets can be normalized using the Dirac product as\footnote{Here, $\overline{\psi}$ is the adjoint of $\psi$.} 
\begin{equation}  \iiint_{-\infty}^{+\infty}\overline{\psi^\mp}\psi^\mp d\alpha d\beta_\pm=1\,,
\end{equation}
thus obtaining 
\begin{equation}
|A^\mp|=1/\sqrt{\pi^3(1+2e^2+2(\bar{k}_+^\mp)^2+2(\bar{k}_-^\mp)^2)}\,.   
\end{equation}

\begin{figure}[ht]
    \centering
\includegraphics[width=0.9\linewidth]{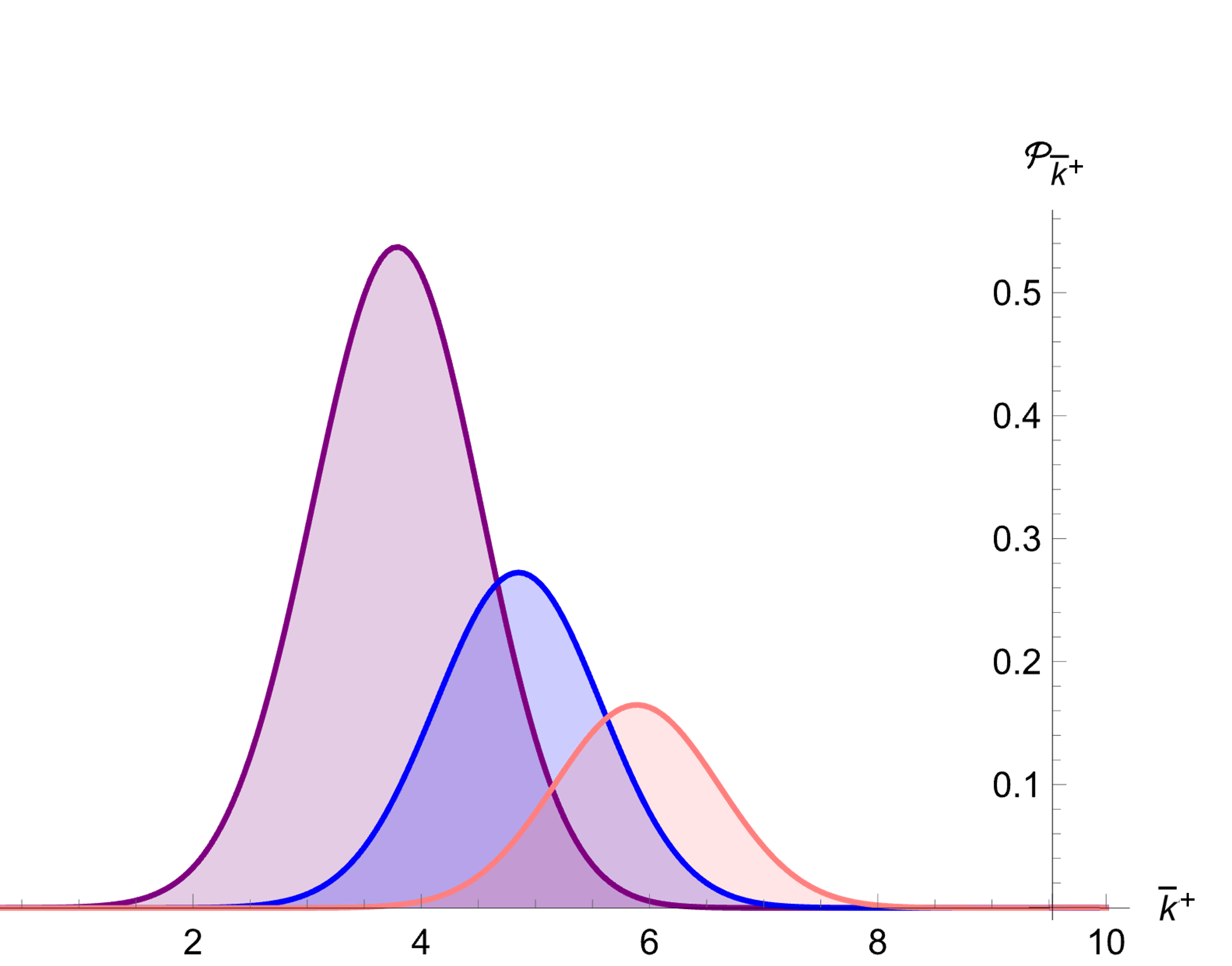}
    \caption{Plot of $\mathcal{P}_{\bar{k}^+}=|S_{\bar{k}^-,\bar{k}^+}|^2$ as function of $\bar{k}^+$ for fixed values of $\bar{k}^-$, where $\bar{k}^-,\bar{k}^+$ are respectively the wavenumbers of the initial (contracting) and final (expanding) state of the Universe.}
    \label{Dirac}
\end{figure}

The same construction as in Section \ref{KG} can also be built in this formalism. In particular, the scattering amplitude from a collapsing to an expanding Universe in the Dirac approach takes the expression
\begin{equation}
S_{\bar{k}^-,\bar{k}^+}=-i\int d\alpha\,d\beta_\pm \, \overline{\psi^-}(\alpha,\beta_\pm)  U(\alpha)\psi^+(\alpha,\beta_\pm)\,,
\label{Diractrans}
\end{equation}
in which $U(\alpha)$ is the considered time-dependent potential. In Figure \ref{Dirac}, we can see the probability density associated when $U(\alpha)$ takes an exponential expression, thus emulating the term in \eqref{BII}. We can notice that the distribution is Gaussian-like and well-defined, demonstrating the validity of this approach. 

The further implementation of this formalism is the analysis of the reflections in the BKL map at a quantum level, by considering the Bianchi II potential as a first step. So far, it is interesting to note that the scalar particle is a good model for the homogeneous Universe at a quantum level, whereas the chaotic behaviour of the anisotropic cosmologies towards the singularity is captured only in a fermionic-like model. 
\section{Discussion and conclusions \label{Conc}}

In this paper, we provided different cosmological applications of the relativistic scattering formalism to homogeneous cosmologies. We started from the isotropic Universe, showing the analogy with a Klein-Gordon framework in which the Universe can be described as a free scalar particle state at a quantum level. Then, we introduced the Quantum Big Bounce approach in which the initial singularity is probabilistically avoided at a quantum level and the semiclassical bouncing dynamics is promoted to a quantum transition from a collapsing to an expanding Universe, thanks to the presence of a time-dependent potential. In particular, we considered a specific expression for the interacting potential, for which we were able to find an exact solution for the Universe wavefunction, in order to compute the Universe volume eigenvalues and search for some imprints of the Quantum Big Bounce at the level of the Wheeler-DeWitt equation. We demonstrated that the Universe volume operator still follows a contracting-like evolution at a quantum level. However, its standard deviation makes the behaviour compatible with an expanding trajectory in the region in which the potential is relevant, thus giving meaning to the probabilistic description of the Quantum Big Bounce. Finally, we promoted the Klein-Gordon formalism to the Dirac one for the Bianchi II model, highlighting how this approach well captures the quantum description of the BKL map. Also in this case, the relativistic scattering approach can be implemented in order to provide a quantum counterpart for the Kasner transitions towards the initial singularity. In all the cases presented in this paper, the probability density associated to the different quantum transitions results well defined and normalizable. 

The most interesting aspect of this analysis is the suitability of the quantum scattering approach to various cosmological scenarios. Furthermore, the Universe shows a scalar-like or fermionic-like behaviour depending on how the anisotropies dynamics is considered. In particular, the chaotic behaviour towards the initial singularity is captured only in the Dirac formalism, whereas the Klein-Gordon one is useful for the description of the Quantum Big Bounce. Moreover, the analysis of the Universe volume operator in the Wheeler-DeWitt formalism demonstrates that the Quantum Big Bounce description is well grounded. In the same spirit, the study of the Kasner transitions with the Dirac scattering formalism can provide a good candidate for a full quantum BKL map, in order to say a definitive word on the fate of chaos in the general cosmological solution.

\begin{description}
    \item[Acknowledgments] The author thanks D. Oriti, C. Rovelli, P. Singh and F. Vidotto for fruitful discussions.
    \item[Funding] This work is partially supported by the MUR PRIN Grant 2020KR4KN2 ``String Theory as a bridge between Gauge Theories and Quantum Gravity'', by the FARE programme (GW-NEXT, CUP:~B84I20000100001), and by the INFN TEONGRAV initiative.
\end{description}

\nocite{*}

\bibliography{bib}

\end{document}